\documentclass[pdflatex,sn-mathphys-num]{sn-jnl}

\usepackage{graphicx}%
\usepackage{multirow}%
\usepackage{amsmath,amssymb,amsfonts}%
\usepackage{amsthm}%
\usepackage{mathrsfs}%
\usepackage[title]{appendix}%
\usepackage{xcolor}%
\usepackage{textcomp}%
\usepackage{manyfoot}%
\usepackage{booktabs}%
\usepackage{algorithm}%
\usepackage{algorithmicx}%
\usepackage{algpseudocode}%
\usepackage{listings}%
\usepackage{bm}
\usepackage{tikz}

\begin{document}

\title{Brightening dark excitons and trions in systems with a Mexican-hat energy dispersion: example of InSe}

\author*{\fnm{Lewis J.} \sur{Burke}}\email{L.J.Burke@lboro.ac.uk}

\author{\fnm{Mark T.} \sur{Greenaway}}\email{M.T.Greenaway@lboro.ac.uk}

\author{\fnm{Joseph J.} \sur{Betouras}}\email{J.Betouras@lboro.ac.uk}

\affil{Department of Physics and Centre for the Science of Materials, Loughborough University, LE11 3TU, United Kingdom}

\abstract{We investigate the properties of momentum-dark excitons and trions formed in two-dimensional (2D) materials that exhibit an inverted Mexican hat-shaped dispersion relation, taking as an example monolayer InSe.
We employ variational techniques to obtain the momentum-dark ground state and bright state (non-zero and zero quasiparticle momenta respectively). These states are particularly relevant due to their peaks in the quasiparticle density of states, where for the momentum-dark ground state, the contribution here is largest due to the presence of a van Hove singularity (VHS).
The momentum-dark systems require a brightening procedure to provide the necessary momentum to become bright. We study the brightening through coupling to phonons and compute the photoluminescence spectrum. This work opens new avenues of research, such as exploiting dark excitons in solar cells and other semiconductor-based optoelectronic devices.}

\maketitle
	\section{Introduction}
	Two-dimensional (2D) van der Waals (vdW) materials offer promising prospects for both fundamental physics research and a wide range of technological applications \cite{2droadmap}. 2D vdW semiconductors with finite bandgaps \cite{miro2014atlas,bhimanapati2015recent}, are particularly promising materials for nanoelectronics and photonics \cite{fiori2014electronics,koppens2014photodetectors}. As a result, understanding these materials' excitonic and optical properties is important as they present new paradigms compared to traditional semiconductors. For example, their weak dielectric screening and strong geometrical confinement give rise to strong Coulomb interaction effects, resulting in rich many-particle phenomena, such as forming a variety of different excitons \cite{Zhang19, tightlybound}, which includes bright and dark excitons.
	
	There are two types of dark excitons (excitons that cannot emit light efficiently due to selection rules): spin-dark, \cite{spin2,spin1, feierabend2020brightening} when the electron and hole have opposite spin \cite{Selig_2018}, and momentum-dark \cite{feierabend2020brightening, madeo2020directly}. A momentum-dark (or momentum-forbidden) exciton is one in which the electron and hole are at different points in momentum-space, such as different valleys in transition metal dichalcogenides (TMDs)\cite{dark_malic}. Here, a photon can provide enough energy to activate a dark state, but it cannot provide the momentum to create a bright excitonic state.
	The dark state energy is thus often lower than their bright counterparts \cite{darklower}. Momentum-dark excitons can be observed experimentally if they couple to phonons \cite{phonon1,phonon2} (and spin-dark excitons if they couple with an in-plane magnetic field \cite{feierabend2020brightening}).
	
	Other interesting many-body excitonic states in vdW materials include interlayer excitons \cite{skinner, inter1} - moir\'e excitons \cite{moire1, guo2020shedding}, biexcitons \cite{biexc} and trions \cite{esser2001theory}.
	A trion is a direct extension of an exciton. This is through adding an extra electron or hole to the system (commonly via electron or hole doping), and consequently, they are negatively or positively charged. This is described as a three-particle complex of a charged exciton \cite{esser2001theory, singh2016trion},  which can be classified as either a singlet or a triplet \cite{singlet_triplet_trion}, depending on the spin-state of the extra particle. If the pair of spins of the electron (negative trion) and the hole (positive trion) are aligned, we have a triplet state and anti-aligned we have a singlet state.
	
	There are other descriptions for a trion which describe it with a four-body Schr\"{o}dinger equation, which includes a conduction band hole in the case of an electron-doped system \cite{rana2020, koksal2021structure}. Dark excitons and trions have longer lifetimes than bright states, with lifetimes roughly of the order of nanoseconds \cite{paylaga2024monolayer,tang2019long}. Therefore, they have potential applications for semiconductor-based optoelectronic devices and solar cell devices for both bright and dark states \cite{mueller2018exciton}.
	
	Experimentally, trion formation in doped systems is realised by electrostatic gating, which can be tuned to allow the formation of positive or negative trions. The resulting exciton and trion photoluminescence (PL) peaks can be measured and analysed \cite{golovynskyi2021trion,GateTunable}. Much of the experimental work has been on TMDs, such as $\text{WSe}_2$ \cite{GateTunable}.  Here, the 2D material is encapsulated by hexagonal boron nitride (hBN) and electrostatic gating is used to control charge and form new bound states. 
	In order to link this experimental work with the theoretical calculations and to study the measured PL and signatures of the trions, recently, a generalised PL formula has been developed that captures both bright and dark trion states via the inclusion of direct and in-direct (phonon-assisted) recombination \cite{perea2024trion}, extending previous work on the excitonic states \cite{Selig_2018,dark_malic,feierabend2020brightening}.
	
	InSe is a 2D semiconductor comprising 4-atom thick monolayers, of covalently bonded Se-In-In-Se atoms, with vdW interlayer bonding. It has shown great potential due to its promising performance in optoelectronic devices and potential for strain engineering and nonlinear optics \cite{ma2013engineering,mudd2013tuning,yuksek2014nonlinear,mudd2016direct,bandurin2017high}.
	An unusual feature of monolayer InSe is the dispersion relation of its valence band, which exhibits a band inversion around the $\Gamma$-point known as a Mexican-hat-shaped dispersion \cite{shubina2019inse}.
	This form of dispersion relation has been shown to give rise to very interesting properties under certain conditions \cite{slizovskiy2015magnetic,slizovskiy2014effect}. In addition, there is a growing interest in systems, such as monolayer InSe, that host Van Hove singularities (VHS) and flattened bands, due to pronounced interaction effects \cite{classen2024high, chandrasekaran2023engineering, chandrasekaran2023practical}. These features make InSe a paradigmatic material for experimental and theoretical investigations \cite{paylaga2024monolayer,shubina2019inse,zultak2020ultra,Falko2020}. To ensure high sample quality, InSe is often encapsulated with hBN, which we will take into account in this work.
	We use a variational approach, based on the 1s hydrogen ground-state wave function, to analyse in detail the exciton physics due to the inverted Mexican-hat present in monolayer InSe. We then investigate the properties of positive and negative trions obtained via hole and electron doping and compare the two and three-quasiparticle ground-state dispersions. The lowest-energy states are momentum-dark. Thus, we also investigate a brightening mechanism via coupling to a phonon to create a virtual bright state to explore the PL spectra.
	
	Our work compares these excitonic and trionic systems. We find strong peaks in the density of states (DOS) of the exciton and trion systems. The peaks at the quasiparticle band minimum correspond to the momentum-dark state, suggesting that the dark states could have a large effect on the optical properties of the system. This is a direct result of the Mexican-hat-shaped valence band dispersion. Our work further reveals that for the electron-doped system, the negatively charged trion at low temperatures dominates the PL intensity. However, as temperature increases, the excitonic intensity becomes more substantial.  In the hole-doped system, which can lead to the formation of a positively charged trion, the momentum bright and dark trions are weakly bound and are comparatively an energetically unfavourable ground-state quasiparticle state of the system compared to that of the exciton.  Our work is easily extended to other 2D materials with Mexican-hat dispersions, such as GaSe.
	
	\section{Results}
	\subsection{Monolayer InSe}\label{InSe-system}
	InSe is an III-VI chalcogenide (metal-group III atom and group VI-chalcogenide),  with a hexagonal structure with unit vectors ${\bm a_1}=\left(a/2,\sqrt{3}a/2\right)$ and ${\bm a_2}=\left(a/2,-\sqrt{3}a/2\right)$, where $a=3.95\textup{~\AA}$ is the lattice constant. The reciprocal primitive lattice vector vectors are  ${\bm b_1}=\left(2\pi/a,2\pi/\sqrt{3}a\right)$ and ${\bm b_2}=\left(2\pi/a,-2\pi/\sqrt{3}a\right)$.
	The high symmetry points $\Gamma$, $K$, $M$ of the hexagonal Brillouin zone (BZ) are: $\Gamma=(0,0)$, $K=\left(4\pi/3a,0\right)$ and  $M=\left(\pi/a,-\pi/\sqrt{3}a\right)$.
	
\begin{figure*}[hbt!]
		\includegraphics[width=\linewidth]{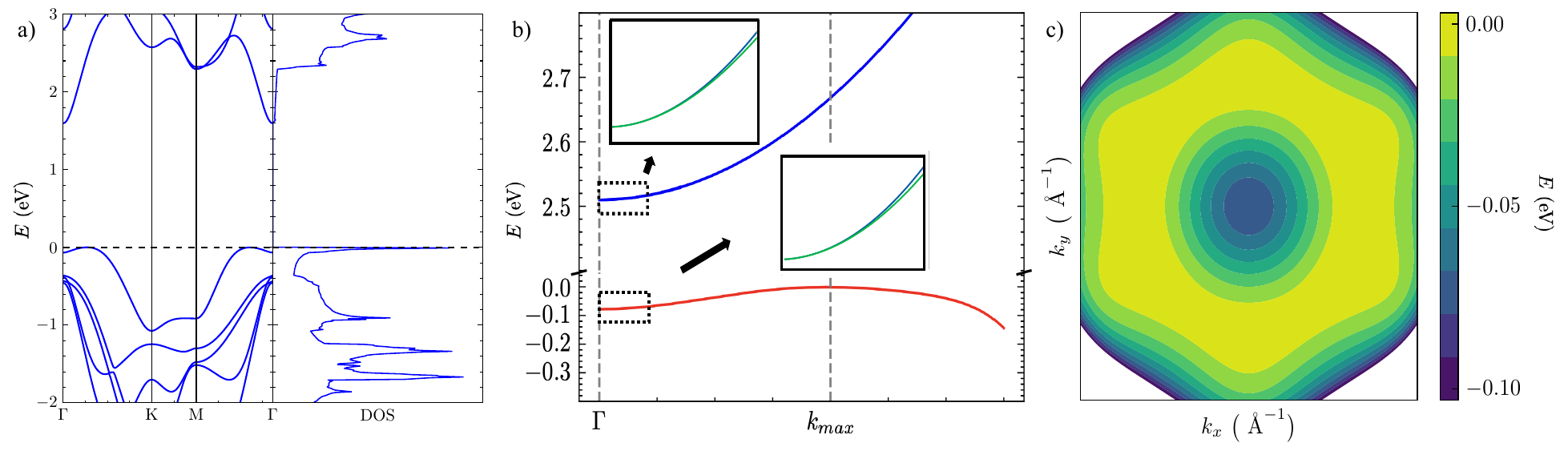}
		\caption{\textbf{Electronic properties of monolayer InSe.}\\
        a) (left) band structure determined by DFT calculations (see Sec. \ref{compdetails}) $E=0$ at valence band maximum, (right) density of states. b) Conduction and valence band determined by the polynomial expression in Eqs. (1-2) c) Colour map of the top-most valence band}
		\label{fig:bands}
	\end{figure*}

	In Fig. \ref{fig:bands}a, we show the band structure and DOS for monolayer InSe obtained using Density Functional Theory (DFT) (see Sec. \ref{compdetails}). It reveals an inverted Mexican-hat shape is present in the top-most valence band at which a VHS is present at the brim of the hat, whereas the conduction band can be modelled by a parabolic expression (see Fig. \ref{fig:bands}b).
	
	DFT  underestimates the size of the bandgap. However, in both excitonic and trionic systems, under the condition of the bandgap being much greater than the binding energy (which implies that we can ignore the overlaps of the electronic states), the binding energy is independent of the bandgap.
	In Fig. \ref{fig:bands}b, near the $\Gamma$-point, the highest valence band and lowest conduction band are modelled by the following polynomial expressions \cite{Falko2016}:
	\begin{equation}
		\begin{split}
			E_c({\bm k})&=\frac{|{\bm k}|^2}{2m_e^*}+E_c^{SO}({\bm k})\\
			E_v({\bm k})&= E_0 +E_1|{\bm k}|^2 +E_2|{\bm k}|^4 +E_3|{\bm k}|^6 \\
			&+E_4|{\bm k}|^6 \cos(6\varphi) +E_5|{\bm k}|^8  +E_v^{SO}({\bm k})
		\end{split}
		\label{eq-conduction/valence}
	\end{equation}
	where the parameters are defined in Table \ref{tab:parameters}.\\
	\begin{table}
		\begin{tabular}{|l|ll|}
			\hline
			$E_0 \hspace{10pt}$ & $-0.078$   & $\text{eV}  $                   \\ \hline
			$E_1\hspace{10pt}$  & $\hspace{7pt}2.915$    & $\text{eV}   \textup{~\AA}^2$   \\ \hline
			$E_2\hspace{10pt}$  & $-38.057$  & $\text{eV}  \textup{~\AA} ^4$ \\ \hline
			$E_3\hspace{10pt}$  & $\hspace{7pt}205.551$  & $\text{eV}    \textup{~\AA}^6$  \\ \hline
			$E_4\hspace{10pt}$  & $\hspace{7pt}3.050$    & $\text{eV}  \textup{~\AA}^6$   \\ \hline
			$E_5\hspace{10pt}$  & $-450.034$ & $\text{eV}   \textup{~\AA} ^8$  \\ \hline
			$m_e^*$ &\hspace{5pt} 0.188 & $m_e$\\ \hline
		\end{tabular}
		\caption{Parameters from \cite{Falko2016}}
		\label{tab:parameters}
	\end{table}
	The DFT calculations show that spin-orbit coupling (SOC)-induced splitting is small, but for completeness and the trion calculations, we include these contributions in Eq. (\ref{eq-conduction/valence}). The terms for the lowest conduction band and highest valence band for the monolayer system, close to the $\Gamma$-point are: \begin{equation}
		\begin{split}
			E_c^{SO}({\bm k})&=\gamma_c s_z |{\bm k}|^3 \cos(3\varphi)\\
			E_v^{SO}({\bm k})&=\gamma_v s_z |{\bm k}|^3 \cos(3\varphi)\\
		\end{split}
		\label{Eq:SOC}
	\end{equation}
	where $\gamma_c=1.49\text{ eV}  \textup{~\AA}^3$ and $\gamma_v=3.11\text{ eV}  \textup{~\AA}^3$ \cite{Falko2016} and $s_z=\pm\frac{1}{2}$.
	This ignores the direct effects of the interaction of the SOC bands, which will be discussed later in this section. 
	
	\subsection{Excitons}\label{exciton-results}
	To obtain the ground-state exciton energy, $E_{ex}(\bm{k})$, in which the electron and hole are at the extrema of the uppermost valence band and lowest conduction band (both of the same spin), we consider the eigenvalue Bethe-Salpeter approach, described in Sec. \ref{method-exciton}. The valence band maximum (VBM), in monolayer InSe, occurs at $\bm{k}={\bm k_{max}} \approx (0.28,0)\textup{~\AA}^{-1}$, which contributes to the degree of how "dark" the system is.
	
\begin{figure}[t!]
		\centering    
		\includegraphics{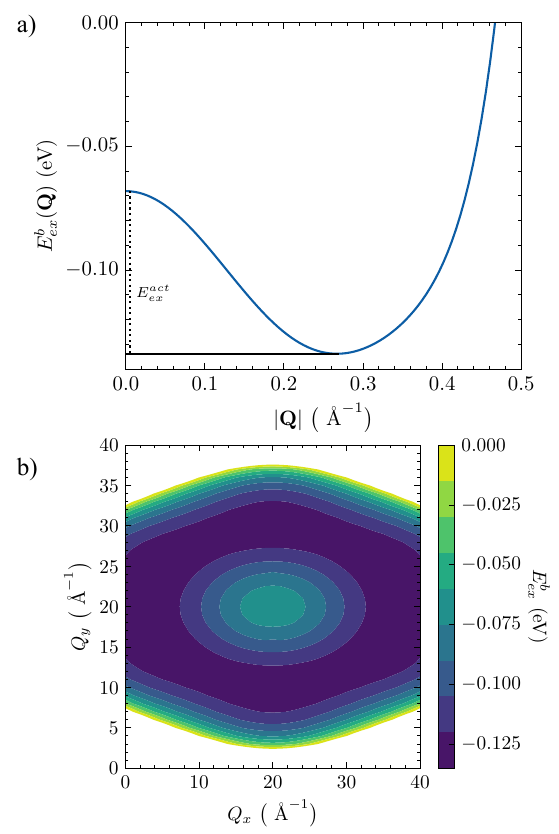}
		\caption{\textbf{Exciton Dispersion}\\
        a) Binding energy, $E^b_{ex}(\bm{Q})$, in the $\Gamma\rightarrow K$ direction. The momentum-dark exciton state is at ${\bm Q_{min}}$. b) Colour map of the exciton dispersion, which reveals the Mexican hat nature over the reduced BZ and the preservation of the hexagonal symmetry.}
		\label{fig:excitonplots}
	\end{figure}

	The binding energy is determined from the calculated exciton energy:\begin{equation}
		E^b_{ex}({\bm Q_{min}})=E_{ex}({\bm Q_{min}})-E_g
	\end{equation}
	where $E_g$ is the the bandgap energy and $\bm{Q}_{min}$ is the wave vector of the lowest exciton state. ${\bm Q_{min}}\neq {\bm 0}$ implies we have a non-zero momentum lowest energy excitonic state. The energy difference between the ${\bm Q}={\bm Q_{min}}$ and ${\bm Q}={\bm 0}$ states gives us the activation energy required to brighten the dark state \cite{Falko2020}
	\begin{equation}
		E^{act}_{ex}=E_{ex}({\bm Q_{min}})-E_{ex}({\bm 0}).
	\end{equation}
	The activation energy thus provides a measure of how distinguishable the bright and dark states would be in an experiment.
	The calculated excitonic binding energy dispersions are shown in Fig. \ref{fig:excitonplots}, where $\bm{Q}$ is the exciton momenta. We find the exciton preserves the hexagonal symmetry of the valence band ($E_4 k^6 \cos(6\varphi)$) and also exhibits a Mexican hat shaped-dispersion. This momentum-dark exciton has a ground-state binding energy and activation energy of:\begin{equation}
		E^{b}_{ex}({\bm Q_{min}}) = -135 \text{ meV}, \;\;\;\;\;
		E^{act}_{ex}=65 \text{ meV}.
		\label{eq-excitonbind}
	\end{equation}
	The variational calculations provide the value for $\beta$ at the finite momentum-state, ${\bm Q_{min}}$, from which we obtain an exciton radius of:
	\begin{equation}
		r_{ex}=\frac{1}{\beta}\approx 17 \textup{~\AA}.
		\label{eq-excitonrad}
	\end{equation}
	we find, $\lambda=1.03\approx1$, confirming that the state is isotropic.
	As this model is appropriate for a Wannier-Mott type exciton and has a relatively large radius, it can be concluded that this is a relatively weakly bound exciton.
	
	\subsection{Trion}\label{trion-results}
	We subsequently consider the energy of the trion quasiparticle following the approach outlined in Sec. \ref{method-trion}. Both positive and negative trions are considered, by adding either an extra hole or an electron to the exciton. We introduce these particles so that they have opposite spin to the ones present in the exciton system, meaning the electron and hole spin-states in their respective trions are anti-aligned, leading to the description of singlet trions (see inset of Fig. \ref{fig:bands}b), which highlights the SOC splitting of the conduction and valence band. This system can be obtained via electron or hole doping (e.g. by the application of gate voltages \cite{GateTunable}) that creates an electron of opposite spin in the bottom-most spin-split conduction band or the annihilation of another electron in the spin-split valence band.
	In Sec. \ref{method-trion}, we describe how to determine the trion energy for the positively and negatively charged trion from which we determine the total binding energy of the three-particle systems,
	\begin{equation}
		\begin{split}
			E^{b,tot}_{tr,-}({\bm Q})&=E_{tr,-}({\bm Q})-2E_g\\
			E^{b,tot}_{tr,+}({\bm Q})&=E_{tr,+}({\bm Q})-E_g
		\end{split}
	\end{equation} 
	we then obtain the trion's binding energy  measured with respect to the exciton's binding energy, this is given as  \cite{rana2020,Rana2016}: 
	\begin{equation}\begin{split}
			E^b_{tr,\pm}({\bm Q})&=E^{b,tot}_{tr,\pm}({\bm Q})-E^b_{ex}({\bm Q})
		\end{split}
		\label{eq-trionbind}
	\end{equation}
	
	\begin{figure}
		\centering
		\includegraphics[width=\textwidth]{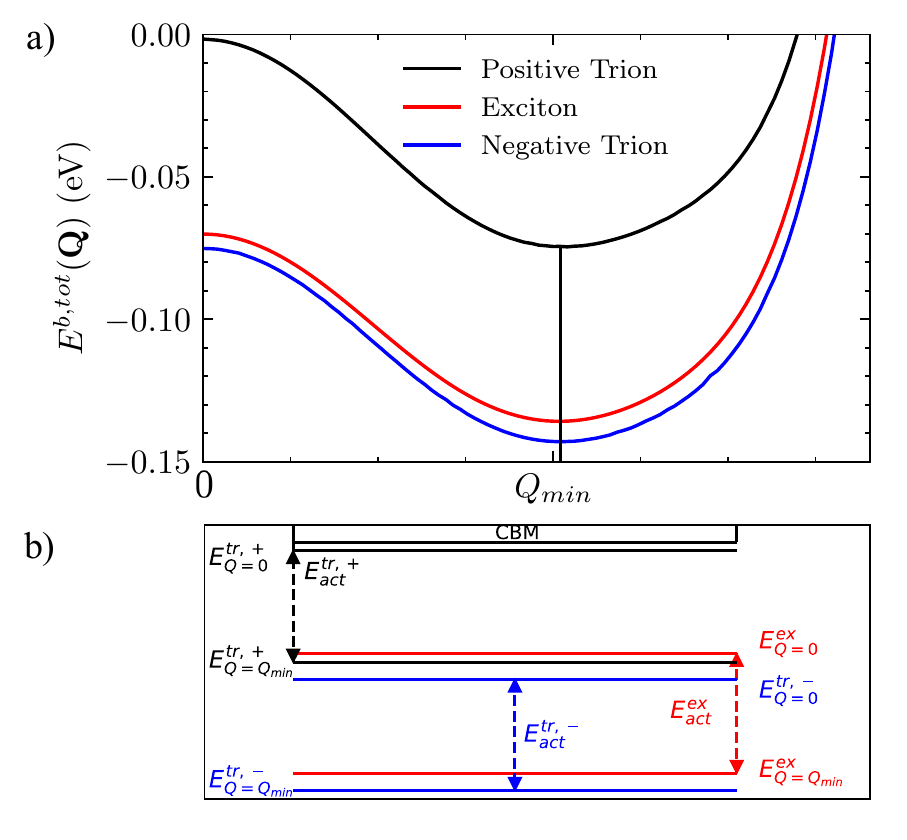}
		\caption{\textbf{Binding energy of the exciton and trion states.}\\
        a) Total three-particle (trion) and two-particle (exciton) binding energies with ${\bm Q_{min}}$ indicated. b) Energy diagram of the bright and dark states for each bound state system with the activation energies indicated with respect to the conduction band minimum (CBM).}
		\label{fig:trionexcompare}
	\end{figure}
	
	$E^b_{tr}(\bm{Q})$ is the energy required for the extra particle to bind to the exciton to form the trion.
	In Fig. \ref{fig:trionexcompare}b, we show the total three-particle binding energies to compare to the excitonic state, the momentum of the minimised state is indicated by ${\bm Q_{min}}$. For the momentum-dark trion at ${\bm Q}={\bm Q_{min}}$ we obtain:
	\begin{equation}
		\begin{split}
			E_{tr,+}^{b}({\bm Q_{min}}) &=\hspace{7pt}65\text{ meV},\;\;\;\;\;\;
			E_{tr,+}^{act} =65 \text{ meV}\\
			E_{tr,-}^{b}({\bm Q_{min}}) &=-10 \text{ meV},\;\;\;\;\;\;
			E_{tr,-}^{act} =65 \text{ meV}
		\end{split}
		\label{eq-trionbindres}
	\end{equation}
	
	Fig. \ref{fig:trionexcompare}a shows that the three-particle negative trion energy system is more strongly bound than both the exciton and the positive trion systems.  The binding energy of the hole to form the positive trion is positive, so it reduces the total binding energy of the three-particle system, this is due to the strong hole-hole interaction at the VHS of the valence band. This result shows that for the hole-doped system, the favourable configuration of the electron and holes is the formation of the exciton-bound state with the addition of a free hole rather than the positively charged trion. This is unlike the electron-doped system which has a negative trion binding energy which leads to a more energetically favourable configuration for the trion system.  
	
	The trion has two length scales given by, radius, $r^{tr}_1$ and $r^{tr}_2$, which are determined by the variational parameters as:
	\begin{equation} 
		r^{tr}_1=\frac{1}{\alpha}\approx 15 \textup{~\AA},\;\;\;\;\;\;
		r^{tr}_2=\frac{1}{\gamma} \approx 50\textup{~\AA}.
		\label{eq-trionrads}
	\end{equation}
	The trion system is characterised by two different  "radii" resulting from the superposition of states in the variational wave function. The values of $r^{tr}_{1,2}$ are effectively the values of the radius of each electron and hole pair comprising the trion. These two radii imply that in the case of the negative (positive) trion, one electron (hole) sits close to the exciton radius, while the other sits further away, which minimises the like-charge repulsive interaction, this is consistent with the behaviour obtained in Ref. \cite{Berkelbach}.
    
	\subsection{Photoluminescence (PL)}\label{results-PL}
\begin{figure}
		\centering
		\includegraphics[width=0.5\textwidth]{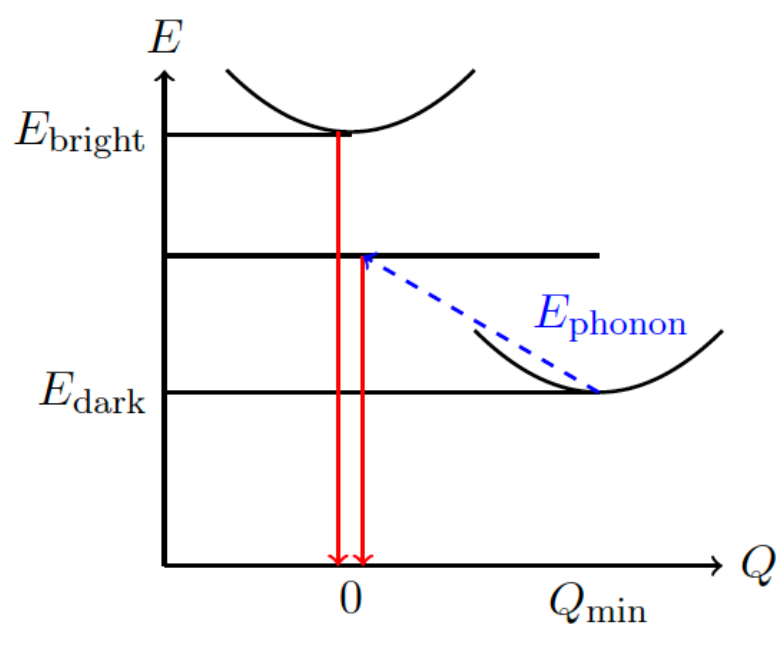}
		\caption{\textbf{Phonon assisted transition}\\
        Bright (${\bm Q}={\bm 0}$) and Dark (${\bm Q}={\bm Q_{min}}$) excitonic states indicated by parabolic dispersion, showing a phonon-assisted transition }
		\label{fig:dark/bright}
	\end{figure}

\begin{figure}
		\centering
		\includegraphics{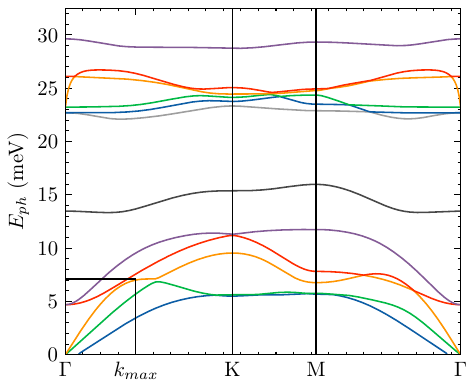}
		\caption{\textbf{Monolayer InSe phonon dispersion.} \\
        Dispersion determined by DFT calculations.}
		\label{fig:ph}
	\end{figure}

\begin{figure}
		\centering
		\includegraphics[width=\textwidth]{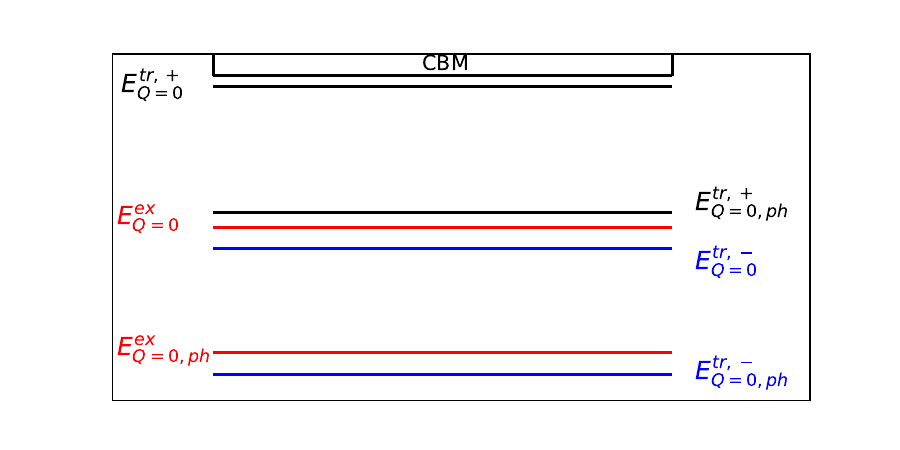}
		\caption{\textbf{Energy diagram of bright and dark states.}\\
        Exciton and trion energies below the conduction band minimum (CBM), including the bright state $(E_{{\bm Q}={\bm 0}})$ and virtual bright state  $(E_{{\bm Q}={\bm 0},ph})$. These energies correspond to peaks in the photoluminescence spectrum; the black/blue lines correspond to the positive/negative trion, and the red lines correspond to the exciton.}
		\label{fig:ex-ph}
	\end{figure}

	Using these excitonic and trionic structures, we study how the bright and dark states (visualised in Fig. \ref{fig:trionexcompare}b)) impact the PL spectrum. The recombination of the electron-hole bound system of the bright state allows the direct emission of a luminescence photon (direct PL). Dark states require phonon coupling to transfer momentum to create a virtual bright state, which leads to indirect PL, see  Fig. \ref{fig:dark/bright}. To account for the phonon coupling, we consider a polaron picture, which can be reduced to a simple phonon scattering process to generate a virtual-bright state. In monolayer InSe, electron-phonon scattering is dominated by longitudinal optical (LO) phonons. Whereas for hole-phonon scattering, the longitudinal acoustic (LA) phonons provide the largest contribution \cite{paylaga2024monolayer,li2019dimensional}. In this work, we consider a single LA phonon that scatters the hole at the valence band maximum, which is at the brim of the Mexican-hat \cite{li2019dimensional}. Given the stronger hole scattering we assume that the emission of a LA phonon with a momentum equal to $\bm{k}_{max}$ and energy $E_{ph}(\bm{k}_{max})$ (see Fig. \ref{fig:ph}) can lead to the brightening of the dark state \cite{paylaga2024monolayer}. In the excitonic picture, this can be visualised as an absorption of the same phonon as described in detail in Sec. \ref{method-phonon}. The coupling to the phonon, subsequently increases the ground-state energy of the dark state, in order to reach the virtual-bright states.
	The result for the exciton virtual-bright state energy due to this change is: \begin{equation}
		E^b_{ex,ph}({\bm Q}={\bm 0})=E^b_{ex}({\bm Q}={\bm Q_{min}})+E_{ph}(\bm{k}_{max})
		\label{eq-exph}
	\end{equation}
	Similarly, for the trion, we require the electron and hole pair of the same spin to be at the same point in momentum-space for both the negative and positive trions thus, the continuation is identical to that of the exciton:\begin{equation}
		E^{b,\pm}_{tr,ph}({\bm Q}={\bm 0})=E^{b}_{tr.\pm}({\bm Q}={\bm Q_{min}})+E_{ph}(\bm{k}_{max}).
		\label{eq-trph}
	\end{equation}
	These states help us populate Fig. \ref{fig:ex-ph}, which is a schematic indication of the bright states and the virtual bright states resulting from the phonon transition.

\begin{figure}
		\centering
		\includegraphics[]{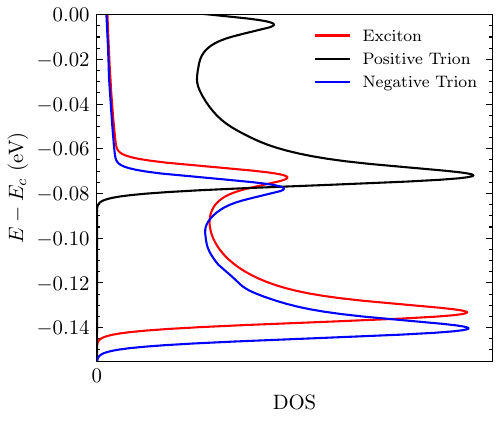}
		\caption{\textbf{Exciton and trion density of states.}\\
        DOS calculated using  Eq. (\ref{eq-dos} corresponding to the trion and exciton dispersion shown in Fig. \ref{fig:trionexcompare}. Two pronounced peaks are observed for each quasiparticle state, the larger peak corresponds to the dark state at which a VHS is present, while the other peak corresponds to the bright state.}
		\label{fig:excitonDOS}
	\end{figure}

    For this calculation, we assume a stationary PL at which thermalisation occurs when the optical excitation takes place.
	In the spectrum, we include the contributions of direct and indirect PL (${\bm Q}={\bm 0}$ and  ${\bm Q}={\bm Q_{min}}$)  for the excitonic \cite{brem2020phonon,feierabend2020brightening} and trion states \cite{perea2024trion}. These states are considered in the calculation due to their large DOS. In Fig. \ref{fig:excitonDOS}, we show the exciton and trion DOS calculated using:\begin{equation}
		g(E)=\sum_{\bm{Q}}\delta(E-E^{b,tot}(\bm{Q}))\approx \frac{1}{\sqrt{2\pi}\sigma}\sum_{\bm{Q}} \text{exp}\left({\frac{-(E-E^{b,tot}(\bm{Q}))^2}{2\sigma^2}}\right).
		\label{eq-dos}
	\end{equation}
	We obtain two distinct peaks corresponding to the dark and bright states. The lowest-energy dark state has a more pronounced peak due to the presence of the VHS in the respective energy dispersion of the quasiparticle DOS. As a result, the PL spectrum will be dominated by these states with large DOS contribution.
    
	The excitonic contribution to the PL spectrum due to the bright (b) and dark (d) (virtual-bright) states are determined using the following expressions \cite{feierabend2020brightening}: 
	\begin{equation}
		\begin{split}
			\text{PL}^{b}(\omega)\propto
			\frac{|M_{{\bm k}}|^2 \gamma_{rad}N_{{\bm Q}={\bm 0}}}{(E_{ex}({\bm Q}={\bm 0})-\omega)^2 +(\gamma_{rad}+\Gamma_{phonon})^2}  
		\end{split}
	\end{equation}
	\begin{equation}
		\begin{split}
			\text{PL}^{d}(\omega)\propto&\frac{|M_{{\bm k}}|^2 }{(E_{ex}({\bm Q}={\bm 0})-\omega)^2 +(\gamma_{rad}+\Gamma_{phonon})^2} \\
			&\times  \frac{|D_Q|^2N^{ex}_{{\bm Q}={\bm Q_{min}}}\eta_{{\bm Q}}\hspace{2pt}\Gamma_{phonon}}{( E_{ex,ph}({\bm Q}={\bm 0})-\omega)^2 +(\Gamma_{phonon})^2}
		\end{split}
	\end{equation}
	where $M_{{\bm k}}$ is the matrix element of the exciton-photon interaction, $N_{{\bm Q}}$ and $\eta_{{\bm Q}}$  are the exciton and phonon occupation factors, $D_{{\bm Q}}$ represents the phonon-exciton interaction matrix and $\gamma_{rad},\Gamma_{phonon}$ are the radiative and phonon-assisted dephasing of the relevant excitonic states.
	
	We consider circularly polarised light $\sigma^{-}$, so that the exciton-photon matrix element is:
	\begin{equation}
		|M_{{\bm k}}|^2=|P_{cv}({\bm k}) \cdot e_{\sigma^{-}}|^2
	\end{equation}
	where $P_{cv}(\bm{k})$ is the interband matrix element which is calculated using the tight-binding model of Ref \cite{Falko2016} so that:\begin{equation}
		\begin{split}
			P_{cv}({\bm k})&=\langle c| \nabla_{\bm k} H|v\rangle\\
			&= \sum_{o,o'} C^\dagger_{c,{\bm k}}(o)C_{v,{\bm k}}(o')\nabla_k\langle o|  H({\bm k})|o'\rangle  
		\end{split}
	\end{equation}
    where the summation is over the orbitals involved in the tight-binding model.
	\begin{figure*}[t!]
		\centering
		\includegraphics[width=\textwidth]{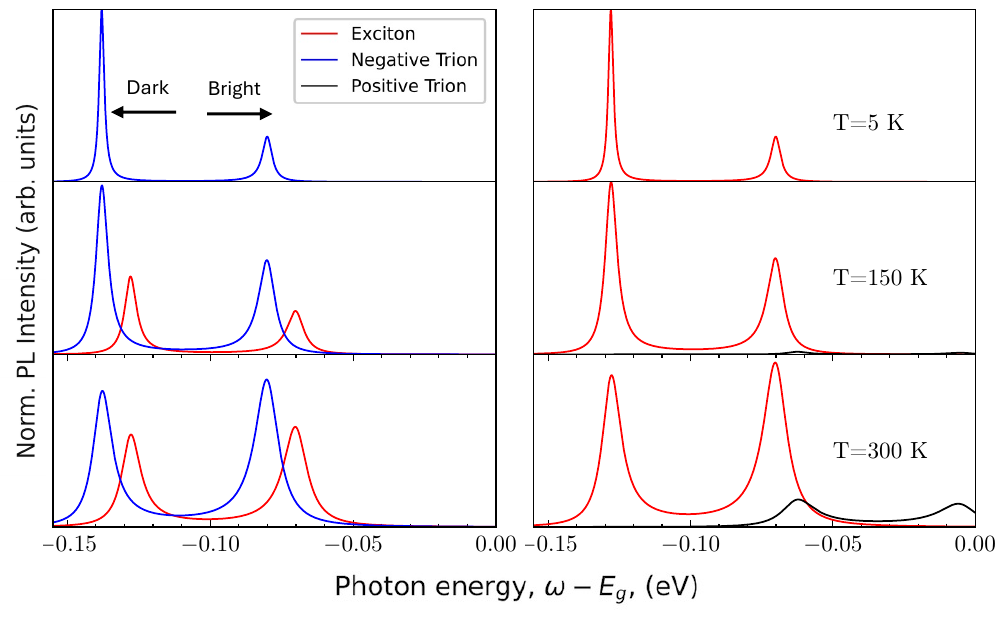}
		\caption{\textbf{PL spectrum for the exciton and trion contributions.}\\ (Left column) compares the exciton (red) and negative trion (blue) in the electron-doped system. (Right column) compares the exciton (red) and positive trion (black) in the hole-doped system. Each row shows how the spectra evolves through temperatures $T=5,150$ and $300$ K. The peak positions slowly change with temperature; this is purely the effect of the dark states. The peaks on the LHS of each pair correspond to the dark (virtual bright) state, and the RHS peak corresponds to the direct PL peak.}
		\label{fig:PLspec}
	\end{figure*} 
The full expression for the PL exciton spectrum can be written as:\begin{equation}
		\begin{split}
			&PL^{ex}(\omega)=\\
			&2\left(\frac{|P_{cv}({\bm k}) \cdot e_{\sigma^{-}}|^2}{(E_{ex}({\bm Q}={\bm 0})-\omega)^2 +(\gamma_{rad}+\Gamma_{phonon})^2} \right)\\
			&\times \bigg[N^{ex}_{{\bm Q}={\bm 0}}\gamma_{rad} + \frac{|D_{\bm{Q}}|^2 N^{ex}_{{\bm Q}={\bm Q_{min}}}\eta_{{\bm Q}}\hspace{2pt}\Gamma_{phonon}}{( E_{ex,ph}({\bm Q}={\bm 0})-\omega)^2 +(\Gamma_{phonon})^2} \bigg]
		\end{split}
		\label{eq-pl_ex}
	\end{equation}
	where the peaks of the spectrum correspond to the photon energy ($\omega$), which matches the bright or virtual-bright state excitonic energy.
	The trion PL spectrum contribution for both systems is given by \cite{perea2024trion}:\begin{equation}
		\begin{split}
			&PL^{tr,\pm}(\omega)\\
			&=2\left( \frac{|M'_{{\bm k}}|^2}{(E^{\pm}_{tr}({{\bm Q}={\bm 0}}) - E^{h/e}_{{\bm k}} -\omega)^2 + (\gamma_{\pm}^{tr,1})^2}\right)\\
			&\times \bigg[N^{tr}_{{\bm Q}={\bm 0}} \gamma^{tr,2}+\frac{N^{tr}_{{\bm Q}={\bm Q_{min}}}|G_{{\bm Q}}|^2 \eta_{{\bm Q}}\hspace{2pt} \gamma_{\pm}^{tr,3}}{( E^{\pm}_{tr,ph}({\bm Q}={\bm 0}) - E^{h/e}_{{\bm k}} -\omega)^2 + (\gamma_{\pm}^{tr,3})^2}\bigg]
		\end{split}
		\label{eq-pl_tr}
	\end{equation}
	with: \begin{equation}
		\begin{split}
			\gamma_{\pm}^{tr,1}&=\gamma^{tr+photon}+\gamma_{\pm}^{tr+phonon}\\
			\gamma^{tr,2}&=\gamma^{tr+photon}\\
			\gamma_{\pm}^{tr,3}&=\gamma_{\pm}^{tr+phonon}
		\end{split}
	\end{equation}
	where we ignore the separate phonon dephasing of the trion due to the electron or hole that remains after the recombination process. Here $M'_{{\bm k}}$ represents the trion-photon matrix element and $|G_{{\bm Q}}|$ represents the trion-phonon interaction matrix. In this case, we assume that the excitonic and trionic phonon couplings and interband transition remain the same. As we assume thermal equilibrium, the exciton and trion occupation factors are well approximated via the Boltzmann distribution 
	    $N^{ex/tr}_{\bm{Q}} \propto \text{exp}\bigg(\frac{-E_{ex/tr}(\bm{Q})}{k_BT}\bigg)$,
 normalized via the partition function; the sum of the Boltzmann contributions of all quasiparticle states (two dark and two bright) in each doped system. For the phonon occupation we use the Bose-Einstein distribution \cite{feierabend2020brightening,perea2024trion}. This is paired with the recoil hole or electron, $E^{h/e}_{{\bm k}}$, \cite{perea2024trion}, which is the energy present to preserve energy and momentum conservation and a result of the recombination process. In this, we use the assumption that all the momenta involved in the recombination process is given to the luminescence photon. Therefore, in these trion systems, we define the recoil energies as the energy of the remaining particle. For the negative trion, the extra electron is situated at the CBM at $\bm{k}=0$. Thus the recoil energy is: $E^e_{{\bm k}}=E_c(\bm{k}=0)=E_g$. For the positive trion, the extra hole is situated at the VBM  at $\bm{k}=\bm{k}_{max}$, which is defined at $E=0$. Thus, the recoil energy is: $E^h_{{\bm k}}=E_v(\bm{k}=\bm{k}_{max})=0$.
	
	In Fig. \ref{fig:PLspec}, we show the PL intensity spectrum for three different temperatures (5, 150, 300 K) for both electron and hole-doped systems. 
For simplicity we assume for the radiative dephasing of the exciton and trion, we use an indicative value of 1 meV, and that the exciton and trion photon couplings are equal. For the phonon-induced dephasing, we adopt the following temperature dependence \cite{perea2024trion}. \begin{equation}
		\begin{split}
			\Gamma_{phonon}&= 1\text{ meV} +(0.01\text{ meV/K})\hspace{5pt}T\\
			\gamma^{tr+phonon}_{-}&=1\text{ meV} +(0.01\text{ meV/K})\hspace{5pt}T\\
			\gamma^{tr+phonon}_{+}&=1\text{ meV} +(0.02\text{ meV/K})\hspace{5pt}T
		\end{split}
		\label{eq-phonondephasing}
	\end{equation}
	
	In the electron-doped system at low temperatures, the negative trion dominates the PL intensity, with the left peak, which corresponds to the dark virtual state, being much more pronounced than the bright peak. This is because the higher energy state has lower occupation, therefore, the recombination process of the trion system is via the dark state and phonon-assisted transitions (as discussed in \cite{brem2020phonon}). With increasing temperature, the bright peak becomes more pronounced than the dark virtual state peak and the contribution from the exciton also increases. At 300 K the trion and exciton peaks have similar intensities. In an experiment, this may be observed as one single broadened peak as seen for systems which exhibit these quasiparticles \cite{christopher2017long,perea2024electrically}.
	This behaviour can be explained by considering two key points: firstly, the dephasing dependence on temperature is similar for both quasiparticles under the assumption of a dominant LA phonon coupling. Secondly, at low temperatures when $k_B T$ is less than the binding energy, the thermal energy is not large enough to destroy the trion state, and the trion bound state would be more energetically favourable due to its larger total binding energy. However, as the thermal energy becomes comparable to or greater than the trion's binding energy, there is a dissociation of the free electron and the exciton (which is still formed due to its higher binding energy). Therefore the trion state becomes less favourable and at 150 K ($\approx13 \text{ meV}$), we see a reduction in the relative strength of the trion peak and its contribution to the PL spectrum.
	
	The temperature dependence of the PL in the hole-doped system (and positive trion) is different.  We find the PL intensity of the positive trion has a much smaller relative intensity compared to the exciton for all temperatures. This is expected since the exciton is the more energetically favourable quasiparticle state. We note that the positive trion will have a much shorter lifetime in comparison to the dominating exciton peaks and would thus be difficult to observe experimentally.
	
	Temperature also plays a role in the energy difference between peaks in the PL spectrum corresponding to the dark and bright states (related to the activation energy in the quasi-particle energy spectrum). This is due to the change in the luminescence photon energy, which corresponds to the dark-virtual state. As temperature is increased, the phonon-induced dephasing term also increases (see Eq. \ref{eq-phonondephasing}), resulting in a small shift of the dark peak so that it moves closer to the peak of the bright state.  Eventually, the two peaks are no longer distinct and they are observed as a single broadened peak. Therefore, as we increase temperature, the difference in photon energy emitted by the bright and dark state peaks effectively reduces.

	\begin{figure}[ht]
		\centering
		\includegraphics[width=10cm, height=8cm]{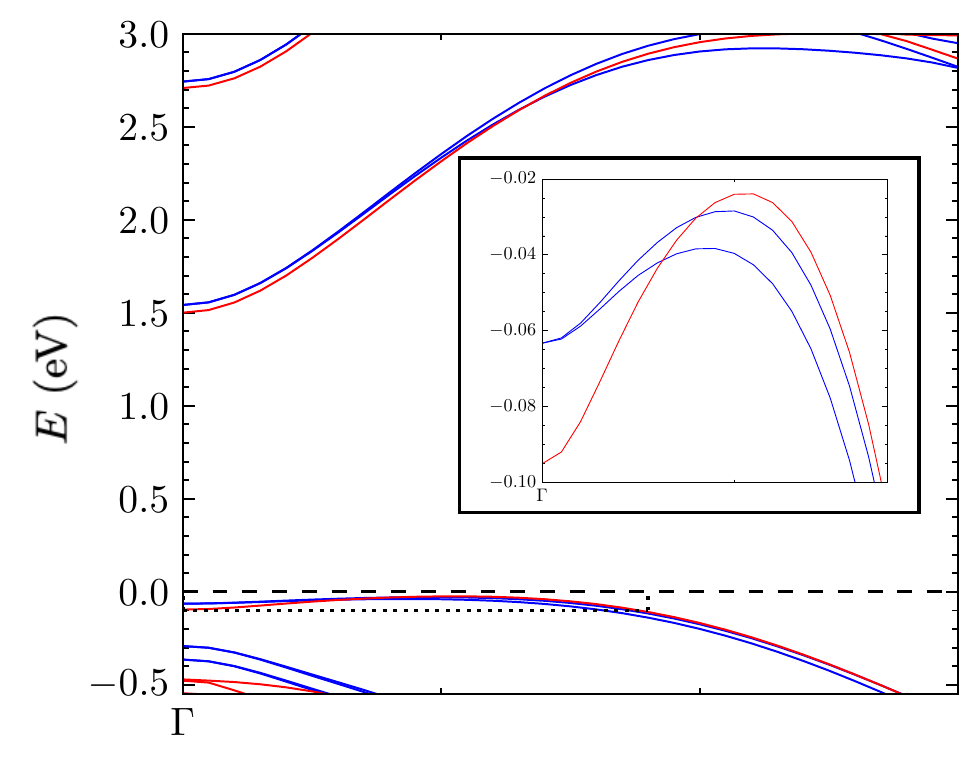}
		\caption{\textbf{Monolayer InSe band dispersion close to the $\Gamma$-point along the $\Gamma-K$ direction.}\\ Blue curve corresponds to DFT calculations with spin orbit coupling included, red is without spin orbit coupling included. Inset is an enlarged section of the Mexican-hat-shaped valence band to highlight the different depths when SOC is included. }
		\label{fig:SOC}
	\end{figure}
	\section{Discussion}
	In this work, we have provided an analysis of the excitonic and trionic spectra of monolayer InSe in which the negative trion has much higher binding energy and pronounced PL intensity compared to the positive trion and the exciton due to larger hole-hole interaction between these states at the VHS in the valence band. The positive trion is less bound than both the exciton and the negative trion. Therefore, it is less energetically favourable in InSe. Further analysis involving more involved methods to solve the BSE equations \cite{zhumagulov2024perspective} can improve quantitatively the picture of these quasiparticles states, beyond the qualitative results that the variational approach provides.
	
	We further show how coupling to phonons can efficiently brighten the momentum-dark states, given the small Mexican-hat depth. This effect allows us to observe the states through photoluminescence.
	Although we have taken as a paradigm the widely studied system InSe, the work can be easily extended to systems with different energy depths (energy difference between $\Gamma$-point and VBM) of the Mexican-hat energy dispersion. For example, GaSe and InTe (and other III-VI metal
	chalcogenides) have been shown to also exhibit an inverted Mexican-hat in the top-most valence band \cite{Falko2014,cai2019synthesis}. To examine the trend in more detail, we consider the effect of the SOC on the shape of the bands of InSe determined via DFT calculations. The inclusion of SOC lifts spin degeneracy over the BZ. However, the effects around the $\Gamma$-point are minimal. In the inset of Fig. \ref{fig:SOC}, one can view the degeneracy lifting, which is not significant in this case. As a consequence, the binding energy is largely unaffected by the inclusion of SOC splitting and the interaction of the bands. A more notable change is that the Mexican-hat dispersion in the top-most valence band is more shallow. The effect of this is that the activation energy of the dark excitonic state is reduced. In the case of systems with a deeper Mexican-hat, the activation energy is subsequently larger, which means that if we assume the bright state has the same binding energy, the resulting darker state will be further away in the PL spectrum. The result of this is that the dark PL peak is identifiable at higher temperatures.
	For accurate quantitative results, as a future work, the derivation and solution of the general Saha equations which describe the detailed balance of the system of photoexcited electronic quasiparticles that forms a multicomponent fluid of excitations in thermodynamic quasiequilibrium need to be solved \cite{manousakis2024excitonic}.
    
	In addition, this work can also account for small strain effects in monolayer InSe; this is because the strain-affected band structure of monolayer InSe is seen as a simple reduction in the bandgap due to a shifting of the conduction band energy \cite{wang2019strain}, which will not affect the result of the binding energies in this model.
	This work further clarifies the effects of constituent particles occupying a flat band and/or systems with a VHS in the examples of the bound quasi-particle structures presented. These findings may stimulate further experimental and theoretical research on optoelectronic and semiconductor-like devices with similar energy dispersion characteristics.
	\section{Methods}
	\subsection{Exciton}\label{method-exciton}
	To find the energy dispersion of the exciton described in monolayer InSe, we solve the Bethe-Salpeter equation (BSE) using the variational method.
	To describe the excitonic system comprising the conduction and valence band, we define the creation operator of the exciton, which acts on the ground-state of the system:
	\begin{equation}
		\Psi_{ex}^\dagger =\frac{1}{\sqrt{A}}\sum_{{\bm k}} \phi({\bm k})c^\dagger_{{\bm k}+{\bm Q}}b_{{\bm k}}|GS\rangle
		\label{eq-exciton creation}
	\end{equation}
	where $c^{\dagger}_{{\bm{k+Q}}}$ is the creation operator of an electron in the conduction band with wave vector ${\bm{k+Q}}$, and $b_{{\bm k}}$ annihilates an electron in the valence band with wave vector ${\bm k}$, creating the corresponding hole.
	
	The Hamiltonian of the monolayer system is given by:
	\begin{equation}H=H_0+H_{int},
		\label{eq-totalH}
	\end{equation}
	where
	\begin{equation}
		H_0=\sum_{{\bm k}} E_c({\bm k})c^\dagger_{{\bm k}} c_{{\bm k}} +\sum_{{\bm k}} E_v({\bm k})b^\dagger_{{\bm k}} b_{{\bm k}} \label{eq-Non-Int}
	\end{equation}
	is the non-interacting Hamiltonian, which describes the kinetic energy of the system.
	\begin{equation}H_{int}=\frac{1}{2A} \sum_{{\bm k_3},{\bm k_4},{\bm q}} V({\bm q})\mathcal{F}({\bm k_3},{\bm k_4},{\bm q})c^\dagger_{{\bm k_4}+{\bm q}}b^\dagger_{{\bm k_3}-{\bm q}}b_{{\bm k_3}}c_{{\bm k_4}}
		\label{eq-Int}
	\end{equation}
	is the interaction Hamiltonian within which $V(\bm{q})$ describes the interaction potential, and $\mathcal{F}$ is the form factor corresponding to the overlap of the electronic states.
	To obtain the eigenvalue equation for the exciton, the standard procedure \cite{methods2d} is to
	first compute the bosonic commutation relation between the Hamiltonian, Eq. (\ref{eq-totalH}), with the exciton creation operator, Eq. (\ref{eq-exciton creation}), and then to compute the fermionic commutation relation with the same quantities. As these commutation relations are equal, this procedure provides the BSE equation \cite{methods2d, Tempelaar_2019}:
	\begin{equation}
		\begin{split}
			E_{ex}\Psi_{ex}^\dagger&= (E_c({\bm k} + {\bm Q})- E_v({\bm k})) \Psi_{ex}^\dagger + \frac{1}{A}\sum_{{\bm k},{\bm q}}\phi({\bm k}-{\bm q})V({\bm q})c^\dagger_{{\bm k}+{\bm Q}}b_{{\bm k}}
			\label{eq-BSE1}
	\end{split}\end{equation}
	where $E_{ex}$ is the exciton energy.
	Using Eq. (\ref{eq-exciton creation}) the above expression can be reduced to \cite{methods2d}:
	\begin{equation}\begin{split}
			E_{ex} \phi({\bm k}) &= [E_c({\bm k}) - E_v({\bm k} - {\bm Q})] \phi({\bm k}) +\frac{1}{A} \sum_{{\bm q}}\phi({\bm k}-{\bm q})V({\bm q})
			\label{eq-BSE2}
	\end{split}\end{equation}
	The form factor, $\mathcal{F}$, in Eq. (\ref{eq-Int}) is set to 1. In principle, the form factor is calculated using the eigenvectors of the electronic states (such as from the tight-binding model) \begin{equation}
		u^\dagger_{{\bm k_3}+{\bm q},c}u_{{\bm k_3},c}u^\dagger_{{\bm k_4}-{\bm q},v}u_{{\bm k_4},v} \label{eq-overlap}
	\end{equation}
	where  $u_{{\bm k},\lambda}$ is the eigenvectors for the $\lambda$ band.
	This approximation leads to the exciton description similar to a Wannier exciton \cite{Falko2020, methods2d}.
	
	The interaction term, $V({\bm q})$, used here is the Fourier transform of the Rytova-Keldysh potential
	\begin{equation}V({\bm q})=\frac{-2\pi e^2}{\sqrt{\kappa_z \kappa_{||}}}\frac{1}{|{\bm q}|(1+r_0|{\bm q}|)},
		\label{Eq-RKpotential}
	\end{equation}
	where $r_0$ is the screening length:
	\begin{equation}r_0=\frac{\sqrt{\varepsilon_z \varepsilon_{||}}-1}{2\sqrt{\kappa_z \kappa_{||}}}d,\end{equation}
	$\varepsilon_z,\varepsilon_{||}$ are the in and out-of-plane permittivity of InSe and $\kappa_z,\kappa_{||}$ are the permittivities of the dielectric environment that encompasses the monolayer InSe structure. In this work, we assume the monolayer is encapsulated by hBN \cite{Falko2020}. Thus, the parameters are \cite{Falko2020}: $\varepsilon_z= 9.5$, $\varepsilon_{||}= 8.6$, $\kappa_z= 6.9$, $\kappa_{||}= 3.7$ and $d=8.32 \textup{~\AA}$ where $d$ represents the InSe layer thickness. 
	To compute the excitonic ground-state through the variational method, we take as a trial wave function the hydrogenic  $1s$ state \cite{methods2d, Rana2014}, in an anisotropic form \cite{skinner, aniso}:
	\begin{equation}\phi_A({\bm k})=\frac{\sqrt{8 \pi/\lambda}\hspace{5pt} \beta^2}{(\beta^2+k_x^2 +k_y^2/\lambda^2 )^{3/2}}
		\label{eq-1s_anisotropic}
	\end{equation}
	Here, $\beta$, which is the inverse radius of the wave function and $\lambda$, which controls the dimensionless anisotropy, are the variational parameters to be determined. This expression is the Fourier transform of the real space wave function $\phi_A({\bm r}) \propto \exp({-\beta\sqrt{r_x^2+(\lambda r_y)^2}})$.
	
	We minimise the expectation value of the excitonic energy, $E_{ex}(\bm{Q})$, via the trial variational wave function $\phi({\bm k})$. To evaluate the integral over the first Brillouin zone, we use the Monkhorst-Pack grid, \cite{monk1, monkhex}, which generates a set of $k$-points from the defined reciprocal lattice. We focus on a region near the $\Gamma$-point where the Mexican-hat is defined.
	
	The expression for the exciton energy $E_{ex}({\bm Q})$ becomes:
	\begin{equation}
		\begin{split}
			&E_{ex}({\bm Q})\sum^{FBZ}_{{\bm k}}\phi_A^*({\bm k})\phi_A({\bm k})\\
			&=\text{min}_{\beta,\lambda}\Bigg(\sum^{FBZ}_{{\bm k}}\Bigg[\phi_A^*({\bm k})\Bigg[[E_c({\bm k}) - E_v({\bm k} - {\bm Q})] \phi_A({\bm k}) +\frac{1}{AN} \sum^{FBZ}_{{\bm k'}} V({\bm k}-{\bm k'})\phi({\bm k'})\Bigg]\Bigg]\Bigg)
			\label{eq-excitonenergy}
		\end{split}
	\end{equation}
	where the summation of ${\bm k'}$ is over the Monkhorst-Pack grid points while keeping ${\bm k}-{\bm k'}$  inside the BZ and contributions from ${\bm k}={\bm k'}$ are ignored.
	\subsection{Trion}\label{method-trion}
	The extension for the ground-state trion follows the exciton, with the addition of the extra hole or electron of the opposite spin to the one previously involved with the exciton. This describes the positive and negative singlet trion systems. Their creation operators are:
	\begin{equation}
		\begin{split}
			\Psi^\dagger_{tr,+}&=\frac{1}{\sqrt{A}}\sum_{{\bm k_1},{\bm k_2}}\phi_{tr}({\bm k_1},{\bm k_2})c^\dagger_{{\bm k_1} + {\bm k_2}+ {\bm Q},\uparrow}b_{{\bm k_1,\uparrow}}b_{{\bm k_2,\downarrow}} \\
			\Psi^\dagger_{tr,-}&=\frac{1}{\sqrt{A}}\sum_{{\bm k_1},{\bm k_2}}\phi_{tr}(\bm {k_1}, {\bm k_2})c^\dagger_{{\bm k_1}+{\bm Q},\uparrow}c^\dagger_{{\bm k_2},\downarrow}b_{{\bm k_1}+{\bm k_2},\uparrow}
		\end{split}
		\label{eq-tr_create}
	\end{equation}
	Taking into account the interaction terms for the three fermions, the eigenvalue equations for the positive and negative trions are:
	\begin{equation}
		\begin{split}
			E_{tr,+}({\bm Q})\phi_{tr}({\bm k_1},{\bm k_2})&=\left[ E_{c}({\bm k_1} +{\bm k_2})-E_{v_1}({\bm k_1}-{\bm Q}) - E_{v_2}({\bm k_2})\right] \phi_{tr}({\bm k_1},{\bm k_2})\\
			&+\sum_{{\bm q}}V({\bm q})\phi_{tr}({\bm k_1}-{\bm q},{\bm k_2}) +\sum_{{\bm q}}V({\bm q})\phi_{tr}({\bm k_1},{\bm k_2}-{\bm q}) \\
			&-\sum_{{\bm q}}V({\bm q})\phi_{tr}({\bm k_1}+{\bm q},{\bm k_2}-{\bm q})\\
			E_{tr,-}({\bm Q})\phi_{tr}({\bm k_1},{\bm k_2})&=[E_{c_1}({\bm k_1}) +E_{c_2}({\bm k_2})
			-E_{v}({\bm k_1}+{\bm k_2}-{\bm Q})]\phi_{tr}({\bm k_1},{\bm k_2})\\
			&+\sum_{{\bm q}}V({\bm q})\phi_{tr}({\bm k_1}-{\bm q},{\bm k_2}) 
			+\sum_{{\bm q}}V({\bm q})\phi_{tr}({\bm k_1},{\bm k_2}-{\bm q})\\
			&-\sum_{{\bm q}}V({\bm q})\phi_{tr}({\bm k_1}+{\bm q},{\bm k_2}-{\bm q})
		\end{split}
		\label{eq-trionenergypos/neg}
	\end{equation}
	where $V({\bm q})$ is the Rytova-Keldysh potential given by Eq. (\ref{Eq-RKpotential}).
	The new variational trial function for the trion has the form \cite{Rana2014,Chang_2021}:
	\begin{equation}\begin{split}
			\phi_{tr}({\bm k_1}&,{\bm k_2})=A(\phi_{\alpha,\beta}({\bm k_1})\phi_{\gamma,\kappa}({\bm k_2})\times(-1)^S \phi_{\alpha,\beta}({\bm k_2})\phi_{\gamma,\kappa}({\bm k_1}) )    
		\end{split}
		\label{Eq-trionwavefn}
	\end{equation}
	It comprises the exciton states previously used (see Eq. (\ref{eq-1s_anisotropic})) for a respective ${\bm k}$ and pair of variational parameters either $(\alpha,\beta)$ or $(\gamma,\kappa)$ in which each pair describes the inverse size of the respective electron-hole pair and the corresponding degree of anisotropy in the system. The factor  $(-1)^S$ determines the nature of the singlet/triplet state.
	We define $\phi_{tr}({\bm k_1},{\bm k_2})$ using Eq. (\ref{Eq-trionwavefn}) with the 1s hydrogen ground-state wave-functions:
	\begin{equation}
		\begin{split}
			\phi^{\text{1s}}_{tr}({\bm k_1},{\bm k_2})&=\Bigg(\frac{1}{(\alpha^2 +k_{1x}^2 +k_{1y}^2/\beta^2 )^{3/2}}\times\frac{1}{(\gamma^2 +k_{2x}^2 +k_{2y}^2/\kappa^2 )^{3/2}}\\
			&+\frac{1}{(\alpha^2 +k_{2x}^2 +k_{2y}^2/\beta^2 )^{3/2}}\times\frac{1}{(\gamma^2 +k_{1x}^2 +k_{1y}^2/\kappa^2 )^{3/2}} \Bigg)
		\end{split}
	\end{equation}
	and employ the same processes used to calculate Eq. (\ref{eq-excitonenergy}) from Eq. (\ref{eq-BSE2}), to obtain the trion energy.
	\subsection{Phonon Coupling}\label{method-phonon}
	To account for the brightening of the dark state described in Sec. \ref{results-PL}, and highlighted in Fig. \ref{fig:dark/bright}, we start by considering a polaron picture. This can be thought of as the combination of electron-phonon and hole-phonon polarons, which make up the excitonic system. The contribution of the phonons to the electron or hole system is then given by the following term \cite{Sio1,Sio2,Sio_2023}:
	\begin{equation}
		\pm\frac{2}{A_{BZ}}\int_{BZ}d{\bm q'}\sum_{\nu}B_\nu({\bm q'})g_\nu^*({\bm k},{\bm q'})A({\bm k}+{\bm q'})
		\label{Eq-polaronphonon}
	\end{equation}
	$\pm$ is the change in energy due to absorption/emission of the phonon, $g$ represents the electron-phonon coupling, $\nu$ is the phonon mode, ${\bm q'}$ is the phonon momentum and $A$, $B$ are envelope functions. $B_\nu$ takes the form \cite{Sio_2023}:\begin{equation}
			B_\nu({\bm q})=\frac{1}{A_{BZ}\hspace{3pt} \hbar \omega_\nu({\bm q'})}\int_{BZ}d{\bm k}\hspace{5pt}A^*({\bm k}+{\bm q'})g_\nu({\bm k},{\bm q'})A({\bm k})
	\end{equation}
	As discussed in Sec. \ref{results-PL}, hole-phonon scattering dominates over electron-phonon scattering. Therefore, we consider only the hole-phonon coupling of the dominate phonon mode, which can be represented as an eigenvalue-type expression for the change in valence band energy, which, in turn, is reduced to a simple scattering term for a single phonon.
	Initially, we have the following:
	\begin{equation}
		\begin{split}
			E_v({\bm k})A({\bm k})&\pm\frac{2}{N}\sum_{({\bm q'})}B({\bm q'})g^*({\bm k},{\bm q'})A({\bm k}+{\bm q'})=E_{v,ph}({\bm k}) A({\bm k})
		\end{split}
		\label{eq-phononcouple}
	\end{equation}
	Taking the expectation value of the expression above to obtain $E_{v,ph}$ we have:
	\begin{equation}
		\begin{split}
			\int &d{\bm k}E_v({\bm k})A^*({\bm k})A({\bm k})\pm\frac{2}{N}\int d{\bm k}A^*({\bm k})\sum_{({\bm q'})}B({\bm q'})g^*({\bm k},{\bm q'})A({\bm k}+{\bm q'})\\
			&=\int d{\bm k}E_{v,p}({\bm k})A^*({\bm k})A({\bm k})
		\end{split}
	\end{equation}
	Here we can see  that:
	\begin{equation}
		\begin{split}
			\frac{1}{A_{BZ}}&\int d{\bm k}A^*({\bm k})g^*({\bm k},{\bm q'})A({\bm k}+{\bm q'})=B^*({\bm q'})\hbar \omega({\bm q'})
		\end{split}
	\end{equation}
	Hence, the term in Eq. (\ref{Eq-polaronphonon}) can be rewritten as:
	\begin{equation}
		\pm \sum_{{\bm q'}}|B({\bm q'})|^2 \hbar\omega({\bm q'})
	\end{equation}
	where $|B({\bm q'})|^2$ represents the average number of phonons involved in the polaron \cite{Sio1,Sio2}. As we are interested in the dominant phonon coupling, which efficiently scatters the hole at the brim of the Mexican-hat to the $\Gamma$-point, we consider the coupling of a single LA phonon at $\bm{q}'=\bm{k_{max}}$ which implies that $|B({\bm q'})|^2=1$ for this particular value of $\bm{q}'$ 
	As a result, we can write:
	\begin{equation}
		\sum_{{\bm q'}}|B({\bm q'})|^2\hbar\omega({\bm q'})= \hbar\omega({\bm k_{max}}) 
		\label{eq-polaron_to_scatter}
	\end{equation}
	this describes the energy of the single phonon at wave-vector ${\bm k_{max}}$,  which is indicated on the phonon dispersion in Fig. \ref{fig:ph}. Therefore, the valence band term simply as:\begin{equation}
			E_v(\bm{k})-\hbar\omega(\bm{k}_{max})=E_{v,ph}(\bm{k}-\bm{k}_{max})
	\end{equation}
		and the non-interacting part of the BSE equation for the exciton which reads as:\begin{equation}
			E_c(\bm{k})-E_v(\bm{k}-\bm{Q}).
		\end{equation} 
		To account for the phonon effects we can rewrite the valence band term, via the emission of the LA phonon\begin{equation}
			\begin{split}
				E_c(\bm{k})&-E_{v,ph}(\bm{k}-\bm{Q}-\bm{k}_{max})\\
				E_c(\bm{k})&-E_v(\bm{k}-\bm{Q})+\hbar\omega(\bm{k}_{max})
			\end{split}
		\end{equation} 
	Therefore, the LA phonon emission, which scatters the hole at the VBM, can be written as the initial excitonic system, which is subsequently transferred to higher energy by  the phonon. This is expected as by the conservation of momentum; the hole-phonon emission implies that for the exciton, we are left with $\bm{Q}+\bm{k}_{max}$, which transferred the minimised exciton state from $\bm{Q}=\bm{Q}_{min}$ to $\bm{Q}=\bm{Q}_{min}+\bm{k}_{max}=\bm{0}$. This implies that in the exciton picture, we show that phonon absorption of the LA phonon leads to the transfer of the dark state to a virtual bright state. Thus, we can write:\begin{equation}
			E_{ex,ph}(\bm{Q=0})=E_{ex}(\bm{Q}=\bm{Q}_{min})+\hbar\omega(\bm{k}_{max})
		\end{equation} 
		The extension for the trion quasiparticles follows identically given the requirement of the spin-aligned electron and holes to be at the same point in momentum-space.
        \begin{equation}
			E^{\pm}_{tr,ph}(\bm{Q=0})=E_{tr,\pm}(\bm{Q}=\bm{Q}_{min})+\hbar\omega(\bm{k}_{max})
		\end{equation} 
        For the full polaron system, one can define the wave function, $A(\bm{k})$, which can be used to form the variational approach encoding the full polaron system, in which one can write the wave function as the 1s-type:\begin{equation}
			A({\bm k})=\frac{1}{(\beta'^2+k_x^2 +k_y^2/\lambda^2 )^{3/2}}.
		\end{equation}
		in which $\beta'$ would now represent the inverse size of the polaron radius, and $\lambda$ describes the anisotropy in the system. This is used for a BSE-type equation which can be variationally optimised as previously discussed. However, given the consideration of the single phonon process we remove the need for this detail in this work.
	
			\subsection{Computational Details}\label{compdetails}
			In this work, the minimisation/variational technique was used in conjunction with the BlackBoxOptim.jl package in the Julia language \cite{blackboxoptim}. The DFT calculations in this paper use the PBE exchange-correlation functional as implemented in Quantum Espresso \cite{QE-2009,QE-2017,doi:10.1063/5.0005082} in which for the band structure calculation (with and without SOC) the cutoffs for the wavefunction and charge density are $90 \text{ Ry}$ and $720 \text{ Ry}$ respectively on a $k$-point grid of $12\times 12 \times 1$, for the phonon calculation a larger $k$-point grid size of $20\times 20\times1$ is used with a  $q$-point grid of $8\times 8\times 1$. 
			\section{Acknowledgements}
			We thank Nicholas Hine and Samuel Magorrian for useful discussions. We are grateful to the UK Engineering and Physical Sciences Research Council for funding via Grant No EP/T034351/1.
			\section{Author Contributions}
			LJB performed the calculations, MTG and JJB initiated and supervised the project. All authors discussed the results and wrote the paper.
			\section{Competing Interests}
			The authors declare no competing interests.
            \section{Data Availability}
Data sets generated during the current study are available from the corresponding author on reasonable request.
            \section{Code Availability}
            The underlying code for this study is not publicly available but may be made
available to qualified researchers on reasonable request from the corresponding
author.
			\bibliography{sn-bibliography.bib}

		\end{document}